\documentclass[twocolumn,aps,prc,tightenlines,floats,floatfix,nofootinbib,preprintnumbers]{revtex4}

\def\st#1{{\kern-1pt} \not\!#1}

\def\sp{\kern +3pt}
\def\sm{\kern -3pt}

\def\spQ{\kern +6pt}

\def\bea{\begin{eqnarray}}
\def\eea{\end{eqnarray}}

\def\sfrac#1#2{{\textstyle \frac{#1}{#2}}}

\newcommand{\bra}[1]{\langle #1|}
\newcommand{\ket}[1]{|#1\rangle}

\def\be{\begin{equation}}
\def\ee{\end{equation}}
\def\ba{\begin{eqnarray}}
\def\ea{\end{eqnarray}}

\usepackage{graphics}
\usepackage{graphicx}
\usepackage{epsf}
\usepackage{amsmath}
\usepackage{amssymb}

\begin{document}

\phantom{0}
\vspace{-0.2in}
\hspace{5.5in}

\preprint{ADP-12-42/T809}

\vspace{-1in}

\title
{\bf Covariant spectator quark model description of the
$\gamma^\ast \Lambda \to \Sigma^0$ transition}
\author{G.~Ramalho$^1$ and  K.~Tsushima$^2$
\vspace{-0.1in}  }

\affiliation{
$^1$CFTP, Instituto Superior T\'ecnico,
Universidade T\'ecnica de Lisboa,
Avenida Rovisco Pais, 1049-001 Lisboa, Portugal
\vspace{-0.15in}
}

\affiliation{
$^2$CSSM, School of Chemistry and Physics,
University of Adelaide, Adelaide SA 5005, Australia
}

\vspace{0.2in}
\date{\today}

\phantom{0}

\begin{abstract}
We study the $\gamma^\ast \Lambda \to \Sigma^0$ transition
form factors by applying the covariant spectator quark model.
Using the parametrization for the baryon core
wave functions as well as for the pion cloud dressing obtained in
a previous work,
we calculate the dependence on the
momentum transfer squared, $Q^2$,
of the electromagnetic transition form factors.
The magnetic form factor is dominated by
the valence quark contributions.
The final result for the transition magnetic moment,
a combination of the quark core and pion cloud effects,
turns out to give a value very close to the data.
The pion cloud contribution, although small, pulls the
final result towards the experimental value
The final result, $\mu_{\Lambda\Sigma^0}= -1.486\, \mu_N$, is about one and 
a half standard deviations from the central value in PDG, 
$\mu_{\Lambda\Sigma^0}= -1.61 \pm 0.08\, \mu_N$. 
Thus, a modest improvement in the statistics of
the experiment would permit the confirmation or rejection of the
present result.
It is also predicted that small but nonzero values for
the electric form factor in the finite $Q^2$ region,
as a consequence of the pion cloud dressing.
\end{abstract}

\vspace*{0.9in}  
\maketitle

\section{Introduction}

There is presently strong motivation to understand
the structure of light baryons in terms of the quark and gluon
dynamics, or quantum chromodynamics (QCD).
Experimentally, however, we have no direct access to the
quarks and gluons.
The experimental studies of the baryon electromagnetic
and weak internal structure is based on
measurements of their form factors.
The measured form factors encode the deviation
of a baryon structure from a pointlike
particle, with the same quantum numbers.
Theoretical studies are often performed
using effective degrees of freedom revealed in the low $Q^2$ region,
such as a baryon core with meson cloud excitations,
where the core is described by constituent quarks
as a first approximation.

Of particular basic interest is
to study the internal electromagnetic structure of the
low-lying spin 1/2 baryons, the octet baryons.
However, except for the proton and neutron,
the electromagnetic structure
of the octet baryons has not yet been uncovered experimentally.
Only magnetic moments of some members of the octet
baryons were measured.
See Ref.~\cite{OctetEMFF} for a detailed bibliography.

Among the octet baryons the reaction $\gamma^\ast \Lambda \to \Sigma^0$
is the only one that is allowed in the electromagnetic transition
between the two different members in the octet under the limit
of isospin symmetry.
The available experimental information
is restricted to the magnitude of the
magnetic moment (at $Q^2=0$):
$|\mu_{\Lambda \Sigma^0}|= 1.61 \pm 0.08 \, \mu_N$,
where $\mu_N$ is the nuclear magneton~\cite{PDG}.
Although the sign of $\mu_{\Lambda \Sigma^0}$
is not known experimentally, $SU(6)$ symmetry suggests that
$\mu_{\Lambda \Sigma^0} = \sfrac{\sqrt{3}}{2} \mu_n \simeq -1.66 \, \mu_N$
\cite{Coleman} ($\mu_n$ is the neutron magnetic moment),
supporting the negative sign.
Estimates based on quark models depend
on the model conventions, and cannot predict
unambiguously the sign.
The implications of the sign will be discussed later.

In this work we study the $\gamma^\ast \Lambda \to \Sigma^0$
reaction using the covariant spectator quark model which was
successfully applied to investigate the electromagnetic
structure of the octet baryons~\cite{OctetEMFF,Medium}.
Using the parametrization determined in Ref.~\cite{Medium}
for the octet baryon electromagnetic form factors,
we predict in this work the electromagnetic
transition form factors for the
$\gamma^\ast \Lambda \to \Sigma^0$ reaction.
We calculate the contributions from the quark core
as well as the pion cloud dressing.
This reaction was studied in the past
using chiral perturbation theory~\cite{Meissner97,Puglia00,Kubis01,Geng,Ahuatzin10},
Skyrme models~\cite{Kunz90,Park92},
chiral quark models~\cite{Kim96,Wagner96,Dahiya03,Cheedket04,Berdnikov07,Sharma10},
other quark models~\cite{Isgur80,Bohm82,Franklin84,Barik,TsushimaCBM,Warns91,VanCauteren04,Franklin02},
QCD sum rules~\cite{Zhu98,Aliev01,Lee11},
lattice QCD~\cite{Leinweber91}, and
some other methods~\cite{Morpurgo,Bartelski05,Jenkins12}.

The $\gamma^\ast \Lambda \to \Sigma^0$
reaction is very interesting to study,
since the initial and final states are different
and have different masses, contrarily
to the case of the elastic reactions of octet baryons.
As a consequence the Dirac-type form factor $F_1(Q^2)$
and the electric form factor $G_E(Q^2)$
vanish at $Q^2=0$, and only the magnetic
form factor $G_M(0)$ survives.
In addition, because of $G_E(0)=0$, we can expect that
the absolute values of $G_E(Q^2)$
is small as a function of $Q^2$
as it happens for the neutron.
This gives an extra interest to study the $Q^2$
dependence of the electric and magnetic form factors.
Another important reason to study the
$\gamma^\ast \Lambda \to \Sigma^0$ transition
is to identify which degree of freedom
gives dominant contributions for the form factors:
the valence quark (quark core) or the quark-antiquark
contribution, namely meson cloud, where
the pion excitations are expected to be dominant.
The covariant spectator quark model, supplemented with the pion cloud dressing,
is therefore particularly convenient
to study the $\gamma^\ast \Lambda \to \Sigma^0$ reaction.

In the covariant spectator quark model,
derived from the covariant spectator theory~\cite{Gross},
a baryon is described as a 3-constituent quark system
where one quark is free to interact with the
photon field, and a pair of non-interacting quarks
is treated as a single on-mass-shell
spectator diquark with an
effective mass $m_D$~\cite{Nucleon,Nucleon2,Omega}.
The quark current is parameterized
based on a vector meson dominance
mechanism as explained in
Refs.~\cite{OctetEMFF,Medium,Nucleon,Omega}.
The model was later improved by the inclusion
of the pion cloud effects~\cite{OctetEMFF,OctetMU,Medium}.
The formalism of the model will be presented in the next section.
The model was also successfully applied to study the
excitation of resonances such as $\Delta$, the Roper,
$N^\ast(1535)$ and others~\cite{NDelta,Delta,Nstar,Delta1600,LambdaS}.

This article is organized as follows:
In Sec.~\ref{secFormFactors} we present
the definitions of the form factors
and explicit expressions of the valence
and pion cloud contributions.
In Sec.~\ref{secResults} we present numerical results.
Finally in Sec.~\ref{secConclusions}, we give summary
and conclusions.

\section{Transition form factors}
\label{secFormFactors}

The $\gamma^\ast \Lambda \to \Sigma^0$
electromagnetic current between
the $\Lambda$ (mass $M_\Lambda$ and momentum $P_+$)
and the $\Sigma^0$ (mass $M_\Sigma$ and momentum $P_-$)
can be represented by
\ba
J_{\Lambda \Sigma}^\mu&=&
\bar u_\Sigma (P_+)
\left\{
F_{1 \Lambda \Sigma}(Q^2) \left( \gamma^\mu - \frac{\not\! q q^\mu}{q^2}\right)
\frac{}{}
\right. \nonumber   \\
& &
\left.
\frac{}{}
+ F_{2 \Lambda \Sigma}(Q^2) \frac{i \sigma^{\mu \nu} q_\nu}{M_\Lambda + M_\Sigma}
\right\} u_\Lambda(P_-),
\label{eqJFF}
\ea
where $F_{1 \Lambda \Sigma}$ and $F_{2 \Lambda \Sigma}$
are respectively
the Dirac-type and Pauli-type form factors.
The sub-index $\Lambda \Sigma$ labels the
reaction to distinguish from
the elastic one.

From Eq.~(\ref{eqJFF}) one can see that
$F_{2 \Lambda \Sigma}(0)$ gives the
transition anomalous magnetic moment in units of
$\sfrac{e}{M_\Lambda + M_\Sigma}$
(analogous to the nucleon case,
$\mu_N= \sfrac{e}{2 M_N}$,
the nuclear magneton with $M_N$ the nucleon mass).
It is therefore convenient to define
the average mass of the initial ($\Lambda$) and
final ($\Sigma^0$) baryon masses,
\ba
M = \sfrac{1}{2}(M_\Lambda+ M_\Sigma).
\ea
We can also define the Sachs form factors for the transition:
\ba
& &
G_{E \Lambda \Sigma}(Q^2)=
F_{1 \Lambda \Sigma}(Q^2) - \tau F_{2 \Lambda \Sigma}(Q^2), \\
& &
G_{M \Lambda \Sigma}(Q^2)=
F_{1 \Lambda \Sigma}(Q^2) + F_{2 \Lambda \Sigma}(Q^2),
\ea
with $\tau= \frac{Q^2}{4M^2}$.
To express $G_{M \Lambda \Sigma}(Q^2)$ in nuclear magneton,
we need to convert by $\sfrac{M_N}{M}G_{M \Lambda \Sigma}(Q^2)$.

\begin{figure}[t]
\includegraphics[width=2.8in]{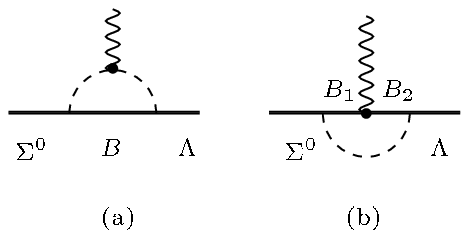}
\caption{\footnotesize
Electromagnetic interaction
within the one-pion loop level (pion cloud) through
the intermediate baryon states $B,B_1$ and $B_2$.
In diagram (a) $B=\Sigma^{\pm}$ while
in diagram (b)
$(B_1,B_2)=(\Sigma^-,\Sigma^-), (\Sigma^+,\Sigma^+)$, or
$(\Lambda,\Sigma^0)$.}
\label{figPionCloud}
\end{figure}

In the covariant spectator quark model the transition
current can be decomposed as
\cite{OctetEMFF,Medium,OctetMU}
\ba
J_{\Lambda \Sigma}^\mu=
Z_{\Lambda \Sigma}
\left[
J_0^\mu + J_\pi^\mu + J_{\gamma}^\mu \right],
\label{eqJ0}
\ea
where $J_0^\mu$ is the current associated
with the direct coupling of photon field to the quark core,
$J_\pi^\mu$ the current resulting from
the photon coupling with the pion [diagram (a)],
and $J_{\gamma}^\mu$ the current related
to the photon interaction with the intermediate baryon
state ($\Lambda$ or $\Sigma$) when one pion is in the air
[diagram (b)].
The factor $Z_{\Lambda \Sigma}$ arises from the
$\Lambda$ and $\Sigma^0$ wave functions
when the pion cloud contributions are included,
to be discussed later.

The final expressions for the form factors can be written as
\ba
& & F_{1 \Lambda \Sigma} (Q^2)=
Z_{\Lambda \Sigma}
\left[
F_{1 \Lambda \Sigma}^0 (Q^2) + \delta F_{1 \Lambda \Sigma} (Q^2)
\right],
\label{eqF1a}
\\
& & F_{2 \Lambda \Sigma} (Q^2)=
Z_{\Lambda \Sigma}
\left[
F_{2 \Lambda \Sigma}^0 (Q^2) + \delta F_{2 \Lambda \Sigma} (Q^2)
\right],
\label{eqF2a}
\ea
where $F_{i\Lambda \Sigma}^0$ ($i=1,2$) are the contributions from
the quark core, and $\delta F_{i \Lambda \Sigma}$ ($i=1,2$)
the contributions resulting from the pion cloud dressing.

\subsection{Valence quark contributions}

In the valence quark sector, we have
\ba
& &
F_{1 \Lambda \Sigma}^0(Q^2)=
\frac{\tau}{1+ \tau} R_{\Lambda \Sigma}(Q^2) {\cal I}(Q^2),
\label{eqF10}\\
& &
F_{2 \Lambda \Sigma}^0(Q^2)= \frac{1}{1+ \tau} R_{\Lambda \Sigma}(Q^2) {\cal I}(Q^2),
\label{eqF20}
\ea
where
\ba
& &
R_{\Lambda \Sigma} (Q^2)= -\frac{1}{\sqrt{3}}
\left\{
f_{1-}(Q^2) + \frac{M}{M_N} f_{2-}(Q^2)
\right\}, \nonumber \\
& & \label{eqGLM2}\\
& &
{\cal I} (Q^2)= \int_k \psi_{\Sigma} (P_+,k) \psi_\Lambda(P_-,k).
\label{eqInt}
\ea
In Eq.~(\ref{eqGLM2})
the functions $f_{i-}$ \mbox{($i=1,2$)} are the isovector
quark form factors defined in Appendix~\ref{appBare}.
(See Refs.~\cite{OctetEMFF,Medium} for details.)
By definition,  $f_{1-}(0)=1$
and $f_{2-}(0)= \kappa_-$,
where $\kappa_-=1.803$ as defined by the
model for the octet baryons in Ref.~\cite{Medium}.
The dependence on the factor $R_{\Lambda \Sigma}$
in Eqs.~(\ref{eqF10}) and (\ref{eqF20})
reflects the isovector character of the reaction.

In Eq.~(\ref{eqInt}) $\psi_\Lambda$ and  $\psi_{\Sigma}$
represent respectively the $\Lambda$ and $\Sigma^0$ radial
wave functions written in a covariant form using
two momentum range parameters,
that characterize the spacial short- and long-range behavior
of the wave functions.
The explicit expressions are given in Appendix~\ref{appBare}.
The symbol $\int_k$ stands for the
covariant integration in the diquark momentum $k$:
$\int_k= \int \frac{d^3 {\bf k}}{(2\pi)^3 2 E_D}$,
where $E_D= \sqrt{m_D^2+ {\bf k}^2}$ is the
diquark energy with the mass $m_D$.

From Eq.~(\ref{eqF10}) we can conclude that $F_{1 \Lambda \Sigma}^0(0)=0$,
as required by the definition of the form factors (\ref{eqJFF}),
when the pion cloud is absent.

So far, the signs of the form
factors are not determined, since the overlap integral~(\ref{eqInt})
depends on the normalization constants of
the $\Sigma^0$ and $\Lambda$ radial wave functions,
$\psi_\Sigma$ and $\psi_\Lambda$, respectively.
However, we have a way to deduce the signs
as to whether the normalization
constants have the same or different
relative sign as will be discussed later.
This relative sign is also important for the other
reactions such as $\gamma^* Y \to \Lambda(1670)$
with $Y=\Lambda, \Sigma^0$ \cite{LambdaS}.
In the following, we start by assuming the same sign for the
both wave function normalization constants.
The choice determines the sign of $F_{2 \Lambda \Sigma}^0(0)$,
since $F_{1 \Lambda \Sigma}^0(0)=0$.
Ultimately, the relative sign of the wave functions can be determined
by experiments for the sign of the transition magnetic moment.

Using Eqs.~(\ref{eqF10}) and~(\ref{eqF20}), we can
calculate also the valence quark contributions for
the electric and magnetic form factors:
\ba
& &
G_{E \Lambda \Sigma}^0(Q^2) =0, \\
& &
G_{M \Lambda \Sigma}^0(Q^2) =
R_{\Lambda \Sigma}(Q^2) {\cal I}(Q^2).
\ea

We can now estimate the valence quark
contribution for the transition magnetic moment.
The result in the equal mass limit is,
$G_{M\Lambda \Sigma}^0(0)=-1.82$,
where $\Sigma^0$ and $\Lambda$ have exactly
the same radial wave functions in their rest frame at $Q^2=0$,
(in this case ${\cal I}(0) =1$).
This leads to $\mu_{\Lambda \Sigma^0} = -1.53\, \mu_N$.

Note however, that we have not yet taken
in consideration the effect of
the pion cloud dressing.
From one side this will require
a redefinition of the
valence quark contribution
by the renormalization factor $Z_{\Lambda \Sigma}$
according to Eqs.~(\ref{eqF1a}) and~(\ref{eqF2a}), and
another side we have additional
contributions for the both form factors
from the pion cloud dressing.

\subsection{Pion cloud contributions}

Now we consider the decomposition (\ref{eqJ0}), and
focus particularly on the contributions $J_\pi^\mu$ and
$J_{\gamma}^\mu$.
Following Refs.~\cite{OctetEMFF,Medium,OctetMU} we
write
\ba
J_{\pi}^\mu &= &
\left(
\tilde B_1 \gamma^\mu + \tilde B_2
\frac{ i \sigma^{\mu \nu} q_\nu}{2 M}
\right) G_{\pi \Lambda \Sigma}, \label{EqJpai} \\
J_{\gamma}^\mu &= &
\left(
\tilde C_1 \gamma^\mu + \tilde C_2
\frac{ i \sigma^{\mu \nu} q_\nu}{2 M}
\right) G_{e \Lambda \Sigma} + \nonumber \\
& &
 \left(
\tilde D_1 \gamma^\mu + \tilde D_2
\frac{ i \sigma^{\mu \nu} q_\nu}{2 M}
\right) G_{\kappa \Lambda \Sigma}, \label{EqJgamma}
\ea
where $G_{\pi \Lambda \Sigma}$ is the
coefficient that includes the coupling
of the photon with the pion,
and $G_{z \Lambda \Sigma}$ $\,(z=e,\kappa)$ are the
coefficients from the couplings
between the photon and the intermediate
baryon states, including the charge coupling $z=e$
and the magnetic coupling $z=\kappa$, where
$e$ and $\kappa$ stand respectively for the charge and anomalous
magnetic moment.
We follow the notations in Refs.~\cite{OctetEMFF,Medium,OctetMU}
except that we need now a double baryon index ($\Lambda \Sigma$)
instead of just $B$, since the initial
and final states are different.
The dependence on the global coupling constant,
$\pi N N$, is included in the coefficients
$\tilde B_i$, $\tilde C_i$ and $\tilde D_i$ ($i=1,2$),
that represent integrals of the corresponding Feynman diagrams,
each as a function of $Q^2$. Here the tilde is a short notation
to remind that they are functions of $Q^2$.
Therefore, the coefficients $G_{\pi \Lambda \Sigma}$ and
$G_{z \Lambda \Sigma}$ depend only on the ratio of the
coupling constants $g_{\pi B B'}/g_{\pi NN}$,
with $g_{\pi NN}^2$ being absorbed in
the respective integral coefficients.
As before \cite{OctetEMFF,Medium,OctetMU},
we assume that the integrals $\tilde B_i, \tilde C_i$
and $\tilde D_i$ ($i=1,2$)
are only weekly dependent on the mass
of the octet baryon members $B$ and therefore
the values of the coefficients hold
for all the octet members.
The expressions for $\tilde B_i, \tilde C_i$
and $\tilde D_i$ ($i=1,2$) are given in
Appendix~\ref{appPionCloud}.

The explicit calculation of $G_{\pi \Lambda \Sigma}$ and
$G_{z \Lambda \Sigma}$ gives
\ba
& &
G_{\pi \Lambda \Sigma} \equiv 0, \\
\label{eqG}
&&
G_{z \Lambda \Sigma} =
- \beta_{\Lambda \Sigma} \left( z_{\Sigma^-} + z_{\Sigma^+}
\right) + \beta_\Lambda z_{\Lambda \Sigma^0},
\ea
with $\beta_{\Lambda \Sigma}= \sqrt{\beta_{\Lambda}} \sqrt{\beta_\Sigma}$,
and
\ba
& &
\beta_{\Lambda} = \frac{4}{3} \alpha^2, \nonumber \\
& &
\beta_{\Sigma} = 4 (1- \alpha)^2,
\ea
where $\alpha \equiv \frac{D}{F+D}$ is defined
in terms of the $SU(3)$ symmetric ($D$) and antisymmetric ($F$)
couplings~\cite{Gaillard84},
and we use the $SU(6)$ quark model value,
$\alpha = 0.6$.
In Eq.~(\ref{eqG}) $z$ stands for
again the charge ($e$) or the
anomalous magnetic moment ($\kappa$) couplings
corresponding to the bare, undressed case.
This means that $z$ is replaced by a function
of $Q^2$ given by $F_{1B}^0(Q^2)$ for $z=e$,
and $F_{2B}^0(Q^2)$ for $z=\kappa$,
where $F_{iB}^0(Q^2)$ $(i=1,2)$ are the bare elastic
form factors\footnote{
In Ref.~\cite{OctetEMFF} we used
$\tilde e_B$ and $\tilde \kappa_B$ to represent respectively
$F_{1B}^0(Q^2)$ and $F_{2B}^0(Q^2)$.}
for the baryon $B$.
In the case of $z_{\Lambda \Sigma^0}$ it
represents the $\gamma^\ast \Lambda \to \Sigma^0$
bare form factors given by Eqs.~(\ref{eqF10}) and~(\ref{eqF20}).
The bare elastic
form factors for $\Sigma^+$ and  $\Sigma^-$
were already obtained in the
previous works~\cite{OctetEMFF,Medium}.

The result $G_{\pi \Lambda \Sigma}=0$ is
a consequence of the cancellation
of the diagrams with a $\Sigma^+$ and a $\Sigma^-$
intermediate states.

Let us now discuss the normalization factor $Z_{\Lambda \Sigma}$,
that is a consequence of the $\Sigma^0$ and $\Lambda$
wave function modification
due to the pion cloud effect.
In the elastic reactions it is easy
to see that $Z_B$ is related to the
charge correction due to the pion cloud~\cite{OctetMU}.
A simple example is the nucleon case.
The correction from the pion cloud
to the proton charge is $3 B_1$  with $B_1=\tilde B_1(0)$.
In this case we can write $G_E(0)= Z_N(1 + 3B_1)$,
where ``1'' corresponds to the proton bare charge.
The correct, dressed charge $G_E(0)=1$, is ensured
by setting $Z_N= 1/(1+ 3 B_1)$.

In general, we can relate the normalization factor $Z_B$
with the derivative of the baryon self-energy~\cite{OctetMU}.
This feature also appears in the cloudy bag model (CBM)~\cite{Thomas}.
The results for the octet baryons $Z_B$
were obtained in Refs.~\cite{OctetMU,OctetEMFF,Medium}.
In particular, for $\Lambda$ and $\Sigma$, we have
\ba
& &
Z_\Lambda=
\left[ 1+ 3 \beta_\Lambda B_1\right]^{-1},
\\
& &
Z_\Sigma=
\left[ 1+ (2\beta_\Sigma +\beta_\Lambda) B_1\right]^{-1}.
\ea
For the $\gamma^\ast \Lambda \to \Sigma^0$ transition
we cannot relate $Z_{\Lambda \Sigma}$
with
the form factor $F_{1 \Lambda \Sigma}(Q^2)$ at $Q^2=0$.
In this case, we include a factor $\sqrt{Z_B}$ for
the initial and final state baryons, which leads to the factor,
$Z_{\Lambda \Sigma}= \sqrt{Z_\Lambda Z_\Sigma}$.

\subsection{Total result}

With the results for the
currents $J_\pi^\mu$ and $J_\gamma^\mu$
together with the definition of $J^\mu_{\Lambda \Sigma}$,
we get,
\ba
F_{1 \Lambda \Sigma} &=&
Z_{\Lambda \Sigma}
\left\{
F_{1 \Lambda \Sigma}^0
+ \left[
 \beta_\Lambda F_{1 \Lambda \Sigma}^0 -
\beta_{\Lambda \Sigma} \left( F_{1 \Sigma^-}^0 + F_{1 \Sigma^+}^0
\right) \right]
\tilde C_1  \right. \nonumber \\
& &
\left.
+
\left[
 \beta_\Lambda F_{2 \Lambda \Sigma}^0 -
\beta_{\Lambda \Sigma} \left( F_{2 \Sigma^-}^0 + F_{2 \Sigma^+}^0
\right)  \right]
\tilde D_1
\right\},
\label{eqF1}
\\
F_{2 \Lambda \Sigma} &=&
Z_{\Lambda \Sigma}
\left\{
F_{2 \Lambda \Sigma}^0
+ \left[
\beta_\Lambda F_{1 \Lambda \Sigma}^0 -
\beta_{\Lambda \Sigma} \left( F_{1 \Sigma^-}^0 + F_{1 \Sigma^+}^0
\right)
 \right]
\tilde C_2  \right. \nonumber \\
& &
\left.
+
\left[ \beta_\Lambda F_{2 \Lambda \Sigma}^0 -
\beta_{\Lambda \Sigma} \left( F_{2 \Sigma^-}^0 + F_{2 \Sigma^+}^0
\right)  \right]
\tilde D_2
\right\}.
\label{eqF2}
\ea
The expressions above show that, we also need to
know the $\Sigma^-$ and  $\Sigma^+$
form factors to calculate the
$\Lambda$ to $\Sigma^0$ transition form factors.
As mentioned already, the $\Sigma^-$ and  $\Sigma^+$
form factors were evaluated in Ref.~\cite{Medium}
in the same framework.
For completeness, we give their
expressions in Appendix~\ref{appBare}.

Note that, in Eq.~(\ref{eqF1})
$\Lambda-\Sigma^0$ contributions from the quark core
vanishes for $Q^2=0$  [$F^0_{1\Lambda \Sigma^0}(0)=0$],
and the same is true for
the terms with $\Sigma^+$ and $\Sigma^-$
[$F^0_{1\Sigma^+}(0)+F^0_{1\Sigma^-}(0) =0$],
and for the Pauli-type form factor contributions [$\tilde D_1(0)=0$].
Therefore, the pion cloud contribution for $F_{1\Lambda \Sigma}(Q^2)$
vanishes at $Q^2=0$, the same as for the bare contributions.
As a consequence $F_{1\Lambda \Sigma}(0)=0$, as expected.

We now calculate the transition magnetic moment given
by $G_{M \Lambda \Sigma}(0) \equiv F_{2 \Lambda \Sigma}(0)$.
From  Eq.~(\ref{eqF2}), we have
\ba
F_{2 \Lambda \Sigma}(0) &=&
Z_{\Lambda \Sigma}
\left\{
\kappa_{0 \Lambda \Sigma}  +
\right.
\nonumber \\
& &\hspace{-1em}
\left.
\left[
\beta_\Lambda \kappa_{0\Lambda \Sigma} -
\beta_{\Lambda \Sigma} \left( \kappa_{0\Sigma^-} + \kappa_{0\Sigma^+}
\right)
 \right]
D_2
\right\},
\ea
where $\kappa_{0B}$ and $\kappa_{0 \Lambda \Sigma}$
are the bare anomalous magnetic moments,
and $D_2$ is the value of $\tilde D_2$ at $Q^2=0$.
With the results $\kappa_{0 \Lambda \Sigma}=-1.817$
(determined before),
and
$\kappa_{0 \Sigma^+}=2.137$,
$\kappa_{0 \Sigma^-}=-0.249$
and
$D_2=0.0821$,
(from Ref.~\cite{Medium}) and
$Z_{\Lambda \Sigma}=0.9246$,
we obtain $G_{M\Lambda \Sigma}(0)= -1.826$,
or $\mu_{\Lambda \Sigma^0}=-1.486\, \mu_N$.
Here the pion cloud contribution is
$-0.12\,\mu_N$.
Comparing the magnitude
of the experimental value of $- 1.61\pm 0.08\, \mu_N$,
our result differs 0.12$\,\mu_N$ from
the central value.
The deviation is almost within the range of error bars.

\section{Results}
\label{secResults}

\begin{figure}[t]
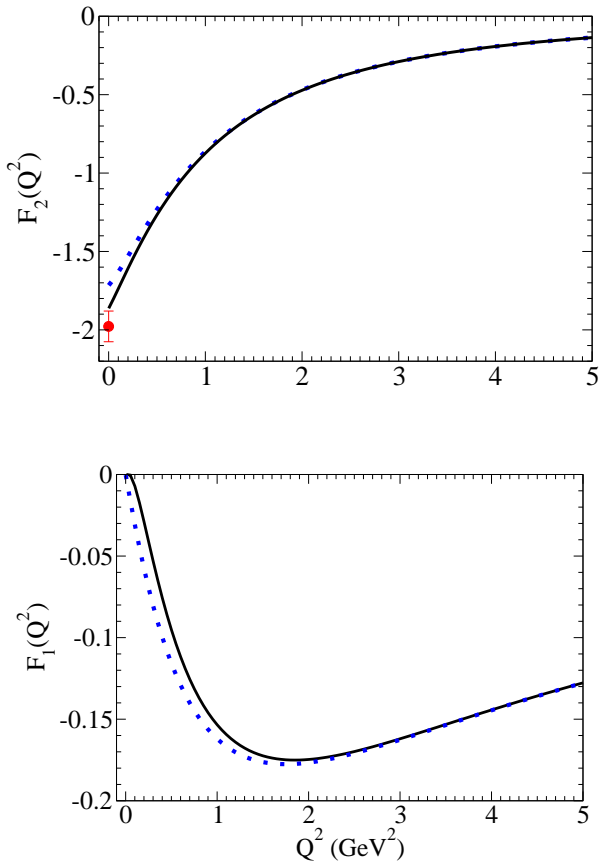

\vspace{.3cm}
\centerline{
\includegraphics[width=3.1in]{F2_mod1} }
\vspace{1.cm}
\centerline{
\includegraphics[width=3.0in]{F1_mod1} }
\caption{\footnotesize{ Dirac- and Pauli-type form factors.
The total and quark core contributions are shown by the solid and dotted lines,
respectively.
}}
\label{figF1F2}
\end{figure}

\begin{figure}[t]
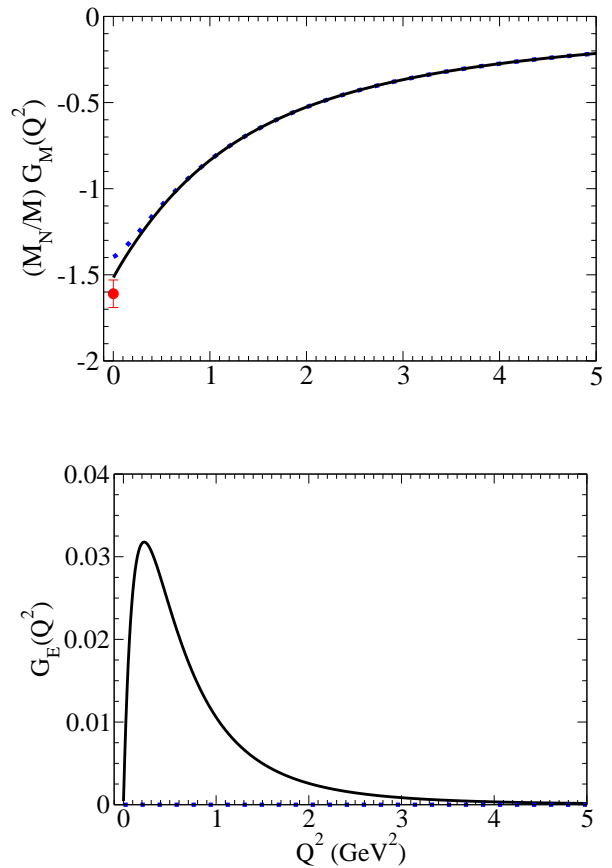

\vspace{.3cm}
\centerline{
\includegraphics[width=3.1in]{GM_mod1} }
\vspace{1.cm}
\centerline{
\includegraphics[width=3.0in]{GE_mod1} }
\caption{\footnotesize{Electric and magnetic form factors.
The total and quark core contributions are shown by the solid and dotted lines,
respectively.
}}
\label{figGEGM}
\end{figure}

The results for the Dirac- ($F_1\equiv F_{1\Lambda \Sigma}$)
and Pauli-type ($F_2 \equiv F_{2 \Lambda \Sigma}$)
form factors, and also the Sachs form factors, are
respectively presented in Figs.~\ref{figF1F2} and~\ref{figGEGM}.
In both figures the solid lines give the
final results from Eqs.~(\ref{eqF1}) and~(\ref{eqF2})
including the pion cloud effects, while
the dotted lines give the contributions
from the quark core (setting $\tilde C_1=\tilde D_1=\tilde C_2=\tilde D_2=0$).
The calculations are performed
including the $\Lambda-\Sigma$ mass difference,
although the approximation,
$M_\Lambda = M_\Sigma= M$, leads to a small
deviation of $\sim 0.5\%$.

In Fig.~\ref{figF1F2} one can see
that the pion cloud effects
(difference between the solid and dotted lines) 
are small but lead the total contribution 
in the direction of the experimental value
$F_2(0) \equiv  F_{2 \Lambda \Sigma}(0)=-1.98\pm0.10$,
extracted from $\mu_{\Lambda \Sigma^0}$.

As for $G_E \equiv G_{E \Lambda \Sigma}$ in Fig.~\ref{figGEGM},
only the total result is visible because the quark core
contribution vanishes.
Therefore, $G_{E \Lambda \Sigma}$ is determined exclusively
by the pion cloud effects.
We note however, based on the scale presented,
that the electric form factor is small
and about 1/3 of the neutron electric form factor
in the similar $Q^2$ range.
As for the magnetic form factor $G_M \equiv G_{M \Lambda \Sigma}$,
corrected by the factor $\sfrac{M_N}{M}$ to be compared with
$\mu_{\Lambda \Sigma^0}$ in Fig.~\ref{figGEGM},
it is dominated by the valence quark contributions.

For $F_2$ and $G_M$ we can observe the fast
falloff of the pion cloud effects.
For $F_1$ and $G_E$ the falloff is slower and has a larger
finite range although the magnitude of the form
factors differs from $F_2$ and $G_M$ by an
order of magnitude.

The results presented here show that we
can study the $\gamma^\ast \Lambda \to \Sigma^0$ reaction
once we know the
$\Lambda$ and $\Sigma^0$ systems.
It is very important to study the $Q^2$ dependence of the
$\Lambda- \Sigma^0$ transition form factors,
since most of the studies were restricted
to the point $Q^2=0$.
Some exceptions are Refs.~\cite{Kubis01,VanCauteren04}.
Our result for $\mu_{\Lambda \Sigma^0} = -1.49 \mu_N$
can well be compared with the other estimates,
$1.4 \mu_N < |\mu_{\Lambda \Sigma^0}|
< 1.9 \mu_N$~\cite{Meissner97,Puglia00,Kubis01,Geng,Ahuatzin10,Kunz90,Park92,Kim96,Wagner96,Dahiya03,Berdnikov07,Sharma10,Cheedket04,Isgur80,Bohm82,Franklin84,Barik,TsushimaCBM,Warns91,VanCauteren04,Franklin02,Zhu98,Aliev01,Lee11,Morpurgo,Bartelski05,Jenkins12},
and is also close to the $SU(6)$ result of
$\mu_{\Lambda \Sigma^0} \simeq -1.66 \mu_N$.
In addition, the available result from lattice QCD
in the quenched approximation is
$\mu_{\Lambda\Sigma^0}=-1.15(16)\, \mu_N$~\cite{Leinweber91},
and underestimates the experimental result.


We note that some other works have
a different sign for $\mu_{\Lambda \Sigma^0}$.
The difference in sign can be a consequence
of the convention of the relative sign between
the $\Lambda$ and $\Sigma^0$ wave functions
(in particular the $\Lambda$ and $\Sigma^0$ quark flavor states).
In our model the contribution from
the quark core (negative) has the same
sign as the pion cloud contribution and is additive.

The relative phase between the $\Lambda$ and $\Sigma^0$
states, that we call $\eta_{\Lambda \Sigma^0}$,
is very important also for other reactions.
So far in the discussion, we have assumed that
$\eta_{\Lambda \Sigma^0}= +1$.
In Ref.~\cite{LambdaS} it is suggested
that the study of the reactions
$\gamma^\ast Y \to \Lambda(1670)$ $(Y= \Lambda,\Sigma^0)$,
can also be used to {\it deduce}
the phase $\eta_{\Lambda \Sigma^0}$.
In this case the results for the form factors are
dependent on the two phases:
$\eta_{\Lambda \Sigma^0}$, and
$\eta_{\Lambda \Lambda(1670)}$, the latter is the relative sign
between the $\Lambda$ and $\Lambda(1670)$ wave functions.
Once the phase $\eta_{\Lambda \Lambda(1670)}$
is determined by one of the reactions,
for instance the case $Y=\Lambda$,
$\eta_{\Lambda \Sigma^0}$ can be fixed by the
other reaction.
In the case $\eta_{\Lambda \Sigma^0}=- \eta_{\Lambda \Lambda(1670)}$,
the suppression of the Pauli-type form factor is expected
in the reaction with $Y=\Sigma^0$~\cite{LambdaS}.

Alternatively, an independent determination
of $\eta_{\Lambda \Sigma^0}$, as in the
reaction $\gamma^\ast \Lambda \to \Sigma^0$,
can be used in the study of the
$\gamma^\ast Y \to \Lambda(1670)$ $(Y= \Lambda,\Sigma^0)$ reactions.

We can also use the present model to calculate
the decay width of $\Sigma^0 \to \gamma \Lambda$,
using our result for  $G_{M \Lambda \Sigma} (0)$.
We obtain $\Gamma = 7.9$ keV, which
is close to the experimental value of
$8.9\pm 0.9$ keV~\cite{PDG}.

\section{Summary and conclusions}
\label{secConclusions}

In this work we have studied the $\gamma^* \Lambda \to \Sigma^0$
transition form factors using the covariant spectator quark model
including the pion cloud effects.
The parameters of the model, including the pion cloud contribution,
are all determined in the previous studies of the electromagnetic
form factors of the octet baryons.

We conclude that
the Dirac- and Pauli-type form factors are dominated
by the valence quark contributions.
However, the relative contributions from the quark core
and the pion cloud, change when we consider the Sachs form factors.
The magnetic dipole form factor is
largely dominated by the valence quark contributions as $F_2$.
As for the electric form factor, the
contributions from the quark core is zero,
therefore $G_E$ is determined exclusively
by the pion cloud contributions.
In all cases the pion cloud effects
fall off faster than those of the valence quarks
with increasing $Q^2$.
We predict that the magnitude of $G_E$ is
at most $\sim 2$\% of that of $G_M$ in the low
$Q^2$ region.
Note that $G_E/G_M$ gives a rough estimate for
the ratio between the pion cloud and valence quark
contributions in our model.
It will be interesting to explore if the proportion
can somehow be accessed by experiment in the near future.

About the transition magnetic moment $\mu_{\Lambda\Sigma}$,
although the magnitude of the pion cloud contribution
is small compared to that of the
valence quark core, it pulls the final result
towards the experimental value.
The final result, $\mu_{\Lambda\Sigma^0}= -1.486\, \mu_N$,
is nearly within the experimental error bars,
about one and a half standard deviations
from the central value in PDG
($\mu_{\Lambda \Sigma^0}= -1.61 \pm 0.08 \, \mu_N$).
Thus, a modest improvement in the statistics of
the experiment would permit the confirmation or rejection of the
present result.

Finally, we recall that the sign of
$\mu_{\Lambda \Sigma^0}$ is given
by the phase $\eta_{\Lambda \Sigma^0}$
between the $\Lambda$ and $\Sigma^0$ wave functions
[$\mu_{\Lambda \Sigma^0} \propto - \eta_{\Lambda \Sigma^0}$].
This phase can also be determined by
experimental measurements of the
$\gamma^* Y \to \Lambda(1670)$ ($Y=\Lambda,\Sigma^0$)
transition form factors.
Thus, the sign of $\mu_{\Lambda \Sigma^0}$ can
be determined consistently with the
$\gamma^* \Lambda \to \Lambda(1670)$
and $\gamma^* \Sigma^0 \to \Lambda(1670)$ reactions.

\vspace{0.2cm}
\noindent
{\bf Acknowledgments:}

G.~R.~would like to thank CSSM at the University of
Adelaide, where part of the work was undertaken,
for the invitation to visit and hospitality.
G.~R. was supported by the Funda\c{c}\~ao para a Cie\^ncia e a
Tecnologia under Grant No.~SFRH/BPD/26886/2006.
This work was also supported partially by the European Union
(HadronPhysics2 project ``Study of Strongly Interacting Matter'')
by the Funda\c{c}\~ao para a Ci\^encia e a Tecnologia,
under Grant No.~PTDC/FIS/113940/2009,
``Hadron Structure with Relativistic Models'',
and (K.~T.~) by the University of Adelaide and
the Australian Research Council through
grant No.~FL0992247 (AWT).

\appendix
\section{Valence quark form factors}
\label{appBare}

\subsection{Wave functions}

Following Ref.~\cite{OctetEMFF}
we consider a generic wave function $\psi_B$
for the octet baryon member $B$
in terms of the baryon momentum $P$
and diquark momentum $k$,
\ba
\Psi_B(P,k)=
\frac{1}{\sqrt{2}}
\left[
\phi_S^0 \ket{M_A}_B +
\phi_S^1 \ket{M_S}_B
\right] \psi_B,
\label{eqPsiB}
\ea
where $\phi_S^{0,1}$ are the spin wave functions,
and $\ket{M_A}_B$ and $\ket{M_S}_B$ are respectively
the mixed-antisymmetric and
mixed-symmetric flavor states with respect to the quarks 1 and 2.
For simplicity, spin and polarization indices are suppressed.

The spin wave functions are expressed by
\ba
& &
\phi_S^0= u_B(P), \\
& &
\phi_S^1=
 - 
\left(\varepsilon_P^{\ast}\right)_\alpha (\lambda) U_B^\alpha(P),
\ea
where $u_B$ is the Dirac spinor,
$\varepsilon_P(\lambda)$ with $\lambda=0,\pm$ is the
diquark polarization state~\cite{Nucleon,FixedAxis}
and $U_B^\alpha$ is the vector spinor given by~\cite{Nucleon,NDelta}
\ba
U_B^\alpha(P)=
\frac{1}{\sqrt{3}} \gamma_5
\left(
\gamma^\alpha - \frac{P^\alpha}{M_B}
\right) u_B(P).
\label{eqUa}
\ea

The flavor wave functions
are listed in Table~\ref{tablePHI}
based on $SU(3)$ symmetry,
which were also given in Ref.~\cite{OctetEMFF}.

The spin and flavor states are expressed
in terms of the states of the quark 3~\cite{Nucleon,Omega,OctetEMFF,OctetMU}.
The total wave function is obtained by
the permutations of the quark states.
However, the present case it
is not explicitly necessary because all
the diquark pairs (12), (23) and (13)
give the equal contributions for the transition current.

\begin{table*}[t]
\begin{center}
\begin{tabular}{l c c c}
\hline
\hline
$B$   & $\ket{M_S}_B$  & &  $\ket{M_A}_B$  \\
\hline
$\Lambda$ &
$\sfrac{1}{2}
\left[ (dsu-usd) + s (du-ud)
\right]$
& &
$\sfrac{1}{\sqrt{12}}
\left[
s (du-ud) - (dsu-usd) -2(du-ud)s
\right]$ \\
$\Sigma^0$ &
$\sfrac{1}{\sqrt{12}}
\left[
s (du+ud) +(dsu+usd) -2(ud+du)s
\right]$
& &
$\sfrac{1}{2}
\left[ (dsu+usd) -s (ud+du)
\right]$ \\
\hline
\hline
\end{tabular}
\end{center}
\caption{Flavor wave functions for $\Lambda$ and $\Sigma^0$.}
\label{tablePHI}
\end{table*}

\subsection{$\gamma^\ast \Lambda \to \Sigma^0$ form factors}

The form factors associated with the photon coupling
with the quarks in the spectator quark model
are calculated by
the relativistic impulse approximation
\cite{OctetEMFF,Medium,Nucleon,Omega},
\ba
J_0^\mu= 3 \sum_{\Gamma} \int_k
\Psi_\Sigma (P_+,k) j_q^\mu
\Psi_\Lambda (P_-,k),
\label{eqJ1}
\ea
where $P_+$ ($P_-$) is the momentum of the final (initial) state,
$\Gamma=\left\{s, \lambda_D \right\}$
holds for the sum in the scalar (spin-0)
and vectorial (spin-1) polarizations
$\lambda_D=0,\pm 1$ of the diquark.
The factor 3 assures the equal contributions
from the 3 independent diquark states (12), (23) and (13).
The quark current operator $j_q^\mu$ can be decomposed as
\ba
j_q^\mu = j_1 \left( \gamma^\mu -\frac{\not \! q q^\mu}{q^2} \right)
+ j_2 \frac{i \sigma^{\mu \nu} q_\mu}{2 M_N},
\label{eqJq}
\ea
where $M_N$ is the nucleon mass.
The quark form factors $j_i$ ($i=1,2$)
act on the third quark state.  
The inclusion of the term  $-{\not\! q} q^\mu / q^2$
in the quark current (\ref{eqJq}) is equivalent
to use the Landau prescription~\cite{Kelly98,Batiz98}
to the final electromagnetic current.
The term restores current conservation but
does not affect the results of the observables~\cite{Kelly98}.

The operators $j_i$ ($i=1,2$) can be decomposed
in the sum of operators in flavor $SU(3)$ space~\cite{Omega,OctetEMFF},
\ba
j_i = \frac{1}{6}
f_{i+} \lambda_0 + \frac{1}{2} f_{i-} \lambda_3
+ \frac{1}{6} f_{i0} \lambda_s, \;\;\;\; (i=1,2),
\label{eqJi}
\ea
where $\lambda_0= \mbox{diag}(1,1,0)$, $\lambda_3=\mbox{diag}(1,-1,0)$
and $\lambda_s= \mbox{diag}(0,0,-2)$.
The operators act on the quark
wave function $q=(uds)^T$ of the third quark.

The functions $f_{i \pm}(Q^2)$ ($i=1,2$) define the quark
electromagnetic form factors.
They are normalized as
$f_{1\,n}(0)=1$ ($n=0,\pm$), $f_{2\pm}(0)=\kappa_\pm$
and $f_{20}(0)=\kappa_s$.
The isoscalar ($\kappa_+$) and isovector ($\kappa_-$)
anomalous magnetic moments are related
with the $u$ and $d$ quark anomalous magnetic moments
by $\kappa_+= 2 \kappa_u - \kappa_d$
and 
$\kappa_-= \sfrac{1}{3}(2 \kappa_u + \kappa_d)$~\cite{Nucleon}.
As for $\kappa_s$, it is the strange quark anomalous
magnetic moment~\cite{Omega}.

The effect of the flavor space overlap
results in the projection of the final and initial states
in the operator $j_i$ ($i=1,2$).
We then define
\ba
& &
j_i^A=  {_\Sigma}\bra{M_A} j_i \ket{M_A}_\Lambda, \nonumber \\
& &
j_i^S=  {_\Sigma}\bra{M_S} j_i \ket{M_S}_\Lambda.
\label{eqJiAS}
\ea

Using the current~(\ref{eqJ1})
with the wave functions defined by Eq.~(\ref{eqPsiB}),
one obtains after some algebra the form factors
defined by Eq.~(\ref{eqJFF}) for the valence quark part:
\ba
& &
\hspace{-1.3cm}
F_{1 \Lambda \Sigma}^0=
\left[
\frac{3}{2} j_1^A
+ \frac{1}{2} \frac{3-\tau}{1+ \tau}
j_1^S
- 2 \frac{\tau}{ 1 + \tau} \frac{M_\Lambda + M_\Sigma}{2 M_N} j_2^S
\right] {\cal I},
\label{eqF10a} \\
& &
\hspace{-1.3cm}
F_{2\Lambda \Sigma}^0=
\left[
\left(
\frac{3}{2} j_2^A
- \frac{1}{2} j_2^S\right) \frac{M_\Lambda +M_\Sigma}{2 M_N}
- 2 \frac{1}{ 1 + \tau}  j_1^S
\right] {\cal I},
\label{eqF20a}
\ea
where ${\cal I}$ is the overlap
integral defined by Eq.~(\ref{eqInt}).

We note that Eqs.~(\ref{eqF10a}) and~(\ref{eqF20a})
are equivalent to the expressions obtained
for the octet baryon electromagnetic
form factors in Refs.~\cite{OctetEMFF,Medium}
if we replace $M_B$ by the average
masses of the $\Lambda$ and $\Sigma^0$,
$M = \sfrac{1}{2}(M_\Lambda + M_\Sigma)$.
This is interesting since
the definitions of the elastic form factors
for the octet baryons and those for the inelastic reaction
$\gamma^\ast \Lambda \to \Sigma^0$ as in Eq.~(\ref{eqJFF}),
are in fact different
[see {\it correction} term in the Dirac form factor].

To get the final results we need only
to use the explicit expressions for
$j_i^A$ and $j_i^S$ for $i=1,2$
given by Eqs.~(\ref{eqJiAS}).
They  can be expressed as,
\ba
j_i^S= -j_i^A= \frac{1}{\sqrt{12}} f_{i-} \;\;\;\;(i=1,2).
\ea
Using these relations, we convert
Eqs.~(\ref{eqF10a}) and~(\ref{eqF20a})
into Eqs.~(\ref{eqF10}) and~(\ref{eqF20}).

\subsection{Elastic form factors}

For the baryons $\Sigma^-$ and $\Sigma^+$
we use the expressions derived
previously~\cite{OctetEMFF,Medium},
\ba
& &
\hspace{-1.3cm}
F_{1 B}^0=
\left[
\frac{3}{2} j_1^A
+ \frac{1}{2} \frac{3-\tau}{1+ \tau}
j_1^S
- 2 \frac{\tau}{ 1 + \tau} \frac{M_B}{M_N} j_2^S
\right] {\cal I}_B,
\label{eqF10B} \\
& &
\hspace{-1.3cm}
F_{2 B}^0=
\left[
\left(
\frac{3}{2} j_2^A
- \frac{1}{2} j_2^S\right) \frac{M_B}{M_N}
- 2 \frac{1}{ 1 + \tau}  j_1^S
\right] {\cal I}_B,
\label{eqF20B}
\ea
where
\ba
{\cal I}_B= \int_k \psi_B(P_+,k) \psi_B(P_-,k).
\ea
The coefficients $j_i^{A,S}$ ($i=1,2$) are presented in Table
\ref{tablePhiB}.

\subsection{Parametrizations for
quark form factors}

To parameterize the quark current, 
we adopt the structure inspired by the vector meson dominance
mechanism as in Refs.~\cite{Nucleon,Omega},
\ba
& &
\hspace{-1cm}
f_{1 \pm} = \lambda_q
+ (1-\lambda_q)
\frac{m_\rho^2}{m_\rho^2+Q^2} + c_\pm \frac{M_h^2 Q^2}{(M_h^2+Q^2)^2},
\nonumber \\
& &
\hspace{-1cm}
f_{1 0} = \lambda_q
+ (1-\lambda_q)
\frac{m_\phi^2}{m_\phi^2+Q^2} + c_0 \frac{M_h^2 Q^2}{(M_h^2+Q^2)^2},
\nonumber \\
& &
\hspace{-1cm}
f_{2 \pm} = \kappa_\pm
\left\{
d_\pm  \frac{m_\rho^2}{m_\rho^2+Q^2} + (1-d_\pm)
\frac{M_h^2 }{M_h^2+Q^2} \right\}, \nonumber \\
& &
\hspace{-1cm}
f_{2 0} = \kappa_s
\left\{
d_0  \frac{m_\phi^2}{m_\phi^2+Q^2} + (1-d_0)
\frac{M_h^2}{M_h^2+Q^2}  \right\},
\label{eqQff}
\ea
where $m_\rho,m_\phi$ and $M_h$ are the masses respectively
corresponding to the light vector meson ($\rho$ meson),
the $\phi$ meson (associated with an $s \bar s$ state),
and an effective heavy meson
with mass $M_h= 2 M_N$ to represent
the short-range phenomenology.
The coefficients $c_0,c_\pm$ and $d_0,d_\pm$ were
determined in the previous studies for the nucleon
(model II)~\cite{Nucleon} and $\Omega^-$~\cite{Omega}.
The values are, respectively,
$c_+= 4.160$, $c_-= 1.160$, $d_+=d_-=-0.686$,
$c_0=4.427$ and $d_0=-1.860$~\cite{Omega}.
The constant $\lambda_q=1.21$ is obtained
so as to reproduce correctly the quark number
density in deep inelastic scattering~\cite{Nucleon}.

\subsection{Parametrizations for radial wave functions}

The radial wave functions for the $\Lambda$ and $\Sigma^0$
(denoted by $B$ below) are defined in terms of the dimensionless
variable~\cite{Nucleon,OctetEMFF,Omega}
\ba
\chi_B= \frac{(M_B-m_D)^2-(P-k)^2}{m_D M_B},
\ea
where $P$ is the baryon momentum and $k$
the diquark momentum.
The radial wave functions are then given by
\ba
\psi_B(P,k)= \frac{N_B}{m_D(\beta_1 + \chi_B)(\beta_3 + \chi_B)},
\ea
where $\beta_i$ ($i=1,3$) are two
parameters that define the momentum scale (in units of $m_D$)
of the radial wave function
in momentum space.
The normalization constant $N_B$
is determined by the condition
$\int_k |\psi_B(\bar P,k)|^2=1$,
where $\bar P=(M_B,0,0,0)$ is the
baryon rest frame momentum.
As in Ref.~\cite{Medium} we use
$\beta_1=0.0532$ and $\beta_3=0.6035$.
In simple terms, $\beta_1$ characterizes
the long-range region in position space
in the radial wave functions,
while $\beta_3$ the short-range.

\begin{table}[t]
\begin{center}
\begin{tabular}{l c c}
\hline
\hline
$B$   & $j_i^S$  &   $j_i^A$  \\
\hline
$\Sigma^+$  & $\sfrac{1}{18}
(f_{i+} + 3 f_{i-} -4 f_{i0}) $ &
 $\sfrac{1}{6} (f_{i+}+3 f_{i-})  $ \\
$\Sigma^-$ & $\sfrac{1}{18}
(f_{i+} - 3 f_{i-} -4 f_{i0}) $ &
        $\sfrac{1}{6} (f_{i+}-  3 f_{i-})           $ \\
\hline
\hline
\end{tabular}
\end{center}
\caption{Mixed-symmetric and mixed-antisymmetric coefficients for
$\Sigma^+$ and $\Sigma^-$
\cite{OctetEMFF}.}
\label{tablePhiB}
\end{table}


\section{Pion cloud dressing}
\label{appPionCloud}

\subsection{Lagrangian}

The relevant part of the Yukawa-type Lagrangian density
for the octet baryons and pseudoscalar
octet mesons without the Dirac structure,
is given by~\cite{Carruthers},
\ba
& &\hspace{-4em}{\cal L}_{PB}
= g_{\pi NN} \bar{N} \vec{\tau} N \cdot \vec{\pi}
\nonumber \\
& &+ g_{\pi\Lambda\Sigma} \left[ \bar{\Lambda} \vec{\Sigma} \cdot \vec{\pi} + H.c. \right]
\nonumber \\
& &+ g_{\pi\Sigma\Sigma} \left[ -i \left( \vec{\bar{\Sigma}} \times \vec{\Sigma} \right) \cdot \vec{\pi} \right],
\label{piSS}\\
& &\hspace{-4em} = g_{\pi NN} \left[
\sqrt{2} \bar{p}n\pi^+ + \sqrt{2}\bar{n}p\pi^-  + (\bar{p}p - \bar{n}n) \pi^0 \right.
\nonumber \\
& &\hspace{-2em} + \sqrt{\beta_\Lambda}\
\left(\ \bar{\Lambda} \Sigma^-\pi^+ + \bar{\Lambda} \Sigma^+\pi^- + \bar{\Lambda} \Sigma^0\pi^0 + H.c. \right)
\nonumber \\
& &\hspace{-2em} + \sqrt{\beta_\Sigma}\ (\ \bar{\Sigma}^0\Sigma^-\pi^+ - \bar{\Sigma}^+\Sigma^0\pi^+
\nonumber\\
& &\hspace{5em} + \bar{\Sigma}^-\Sigma^0\pi^- - \bar{\Sigma}^0\Sigma^+\pi^-
\nonumber\\
& &\hspace{7em} +  \bar{\Sigma}^+\Sigma^+\pi^0 - \bar{\Sigma}^-\Sigma^-\pi^0\ )
\left. \right],
\label{piSSphys}
\ea
where $\beta_\Lambda = 4\alpha^2/3$ and $\beta_{\Sigma} = 4 (1 - \alpha)^2$
with 
$\alpha = \frac{D}{F+D}$.
The $D$ and $F$ correspond respectively
to the $SU(3)$ symmetric and antisymmetric couplings.

\subsection{Pion cloud parametrization functions}
\label{secPionCloud2}

We consider the following
parametrizations for the functions
$\tilde B_i,\tilde C_i$ and $\tilde D_i$
in Eqs.~(\ref{EqJpai}) and~(\ref{EqJgamma}),
\ba
& & \tilde B_1 =
B_1( 1 + t_1 Q^2) \left( \frac{\Lambda_1^2}{\Lambda_1^2 + Q^2}
\right)^5, \label{eqB1}
 \\
& & \tilde C_1= B_1\left(\frac{\Lambda_{1}^2}{\Lambda_{1}^2 + Q^2} \right)^2,
\label{eqC1}
\\
& & \tilde D_1= D_1^\prime \frac{Q^2\Lambda_{1}^4}{(\Lambda_{1}^2 + Q^2)^3},
\\
& & \tilde B_2 =
B_2 (1 + t_2 Q^2)\left( \frac{\Lambda_2^2}{\Lambda_2^2 + Q^2}
\right)^6, \label{eqB2} \\
& & \tilde C_2= C_2\left(\frac{\Lambda_{2}^2}{\Lambda_{2}^2 + Q^2} \right)^3,
\\
& & \tilde D_2= D_2\left(\frac{\Lambda_{2}^2}{\Lambda_{2}^2 + Q^2} \right)^3,
\label{eqD2}
\ea
where
$B_1,B_2,C_2, D_2$ are
constants given respectively by
$\tilde B_1(0),\tilde B_2(0),\tilde C_2(0), \tilde D_2(0)$,
and $\Lambda_1,\Lambda_2$ are two cutoffs, and
$D_1^\prime$ is also a constant
defined by $D_1^\prime = \sfrac{1}{\Lambda_{1}^2}\sfrac{d D_1}{d Q^2}(0)$.
The coefficients $t_1$ and $t_2$ are defined next.
The parametrization for Eqs.~(\ref{eqB1})-(\ref{eqD2})
is chosen to reproduce the charge of the
dressed nucleon (which requires
$\tilde C_1(0) = \tilde B_1(0)$ and
$\tilde D_1(0)=0$)~\cite{OctetEMFF,Medium,OctetMU}
and also to simulate the chiral behavior of the nucleon radii
in the limit of the very small pion mass $m_\pi$~\cite{Medium}.
The coefficients $t_1$ and $t_2$
are obtained as~\cite{Medium},
\ba
& &
t_1= \frac{1}{Z_N B_1}
\left(
\frac{1}{24} 
\frac{5g_A^2+1}{8 \pi^2 F_\pi^2}
\log m_\pi
+ b_1^\prime \right),\\
& &
t_2=
\frac{1}{Z_N B_2}
\left(
-\frac{1}{24} 
\frac{g_A^2}{8 \pi F_\pi^2}
\frac{M_N}{m_\pi}
+ b_2^\prime
\right),
\ea
where $Z_N$ is the nucleon renormalization factor
$Z_N=1/(1+3 B_1)$,
$g_A$ the nucleon axial vector coupling constant,
and $F_\pi$ the pion decay constant.
As for $b_1^\prime$ and $b_2^\prime$
they are two additional parameters determined
by the nucleon isovector radii \cite{Medium}.

All the parameters were determined in Ref.~\cite{Medium}
in the study of the octet baryon electromagnetic form factors,
and the values are listed in Table~\ref{tabPionCloud}.

\begin{table}[b]
\begin{center}
\begin{tabular}{c c c c c }
\hline
\hline
    \sp $B_1,B_2$ \sp \sp & \sp\sp\sp $C_2$ \sp\sp\sp
                          & \sp $D_1^\prime,D_2$ \sp &
\sp $b_1^\prime, b_2^\prime$ \sp
    & \sp $\Lambda_1,\Lambda_2$(GeV)  \\
\hline
      0.0510  &             &     -0.1484    & \sp 1.036 &  0.786   \\
      0.2159  &  0.00286    & \sp  0.08214    &    -1.987 &  1.132   \\
\hline
\hline
\end{tabular}
\end{center}
\caption{
Parameters associated with the pion cloud~\cite{Medium}.
}
\label{tabPionCloud}
\end{table}

\end{document}